\documentclass{aa}
\usepackage{txfonts}
\usepackage{graphicx}

\begin{document}
\title{Interstellar Scintillation of PSR\ J0437$-$4715}

\author{T.V. Smirnova\inst{1,2} \and C.R. Gwinn\inst{3} \and V.I. Shishov\inst{1,2} }

\institute{Pushchino Radio Astronomy Observatory of Lebedev Physical
Institute, 142290 Pushchino, Russia
\email{tania@prao.psn.ru, shishov@prao.psn.ru}
              \and
Isaac Newton Institute of Chile, Pushchino Branch
              \and
Department of Physics, University of California Santa Barbara, CA 93106, USA
\email{cgwinn@physics.ucsb.edu}
              }

\date{Received 00.00.05 / Accepted 00.00.05 }

\abstract
{}
{We studied the turbulence spectrum of the local interstellar plasma in the direction of PSR\ \object{J0437$-$4715},
on the basis of our observations and those reported earlier by others.}
{We combine these data to 
form a structure function for the variations of phase along the line of sight to the pulsar.
For observations that did not report them, we infer modulation indices from a theoretical model.}
{We find that all of the observations 
fit a power-law spectrum of turbulence with index $n=3.46\pm 0.20$. 
We suggest that differences among reported values for scintillation bandwidth and timescale for this pulsar arise from differences in observing parameters.
We suggest that refractive effects dominate for this line of sight,
with refraction angle about twice the diffraction angle at 330\ MHz observing frequency. }
{We suggest that the scattering of this pulsar lies in a layer of enhanced 
turbulence, about 10\ pc from the Sun.
We propose that the flux variations of the
extragalactic source PKS\ 0405$-$385 arise in the same scattering layer.}


\maketitle

\section{Introduction}

Study of the interstellar medium (ISM) for nearby pulsars leads to understanding of the local interstellar medium.
As Shishov et al. (\cite{shi03}) showed, analysis of multi-frequency observations of pulsar scintillation is
critical to understanding the turbulence spectrum of the interstellar plasma. Multifrequency observations
are able to detect details that cannot be seen from observations at a single frequency.  
For example, construction of the
phase structure function from multi-frequency observations of \object{PSR\ B0329$+$54} resulted in the detection of strong angular
refraction along the line of sight to this pulsar. 
This structure function also showed that the power-law index of the spectrum,
$3.50 \pm 0.05$, differs significantly from the Kolmogorov value of 11/3.  
The Kolmogorov spectrum describes much data on pulsar
scintillation quite well, over a large range of  spatial scales 
(Armstrong et al. \cite{arm95}, Shishov \& Smirnova \cite{shi02}), 
but in particular directions,
and particularly along short lines of sight which sample only a small part of the interstellar medium,
the spectrum can differ from the Kolmogorov form.

In this paper we study the turbulence spectrum of plasma along the line of sight to PSR\ J0437$-$4715. This is one
of the closest pulsars.  It has distance $R = 150$\ pc, and its transverse velocity is
100 km/sec (van Straten et al. \cite{Van01}).  
It is quite strong over a wide range of observing frequencies.
We combine the data in the accompanying paper
(Gwinn et al. \cite{paper1}, hereafter Paper 1; also Hirano \cite{hir01}) 
with observations by others, 
and construct the phase structure
function for this pulsar in the time and frequency domains.  
From examination of the structure functions, we conclude that refractive effects are important for this line of sight,
with refraction angles about twice the diffraction angle.
Based on comparison of data for the flux variation of
the extragalactic source PKS\ 0405$-$385 with the scintillation parameters of PSR\ J0437$-$4715, we suggest that that the intensity
variations for both are caused by the same scattering layer, located near the Earth at a distance $\approx 10$\ pc.

\section{Observational Data}

\subsection{Scintillation Parameters}

Observers commonly describe a pulsar's scintillation by its
characteristic scales: scintillation bandwidth
and scintillation timescale.
These describe the decline of the autocorrelation function of the intensity,
with frequency and time lag, from its peak at
zero time and frequency lag.
Observers usually 
normalize 
the autocorrelation function by its peak value.
This is the square of the modulation index, if the data are normalized by 
the mean intensity $\langle I\rangle$. 
The modulation index is: 
$m = \sqrt{ \langle I^2-\langle I\rangle^2\rangle }/\langle I\rangle$.
Scintillation timescale $t_{ISS}$ is the time lag where the
autocorrelation falls to $1/e$ of this central value,
and scintillation bandwidth $\Delta\nu_{ISS}$ is the frequency lag
where it falls to $1/2$ of that value.
For fully sampled data,
in strong scattering where differences among wave paths are many radians,
the modulation index is $m=1$. 
If the data do not span several scintillation bandwidths,
or frequency scales, then the modulation index is less than 1,
as discussed in the Appendix below.

Although the autocorrelation contains a great deal of information on the 
structure function of the density inhomogeneities responsible
for scattering (Shishov et al. \cite{shi03}),
the characteristic scales of scintillation carry a very
limited part of that information.
A more detailed study of the structure function of the
inhomogeneities requires the
modulation index and the form of the autocorrelation function.

\subsection{Summary of Observations for PSR\ J0437$-$4715}

\begin{figure}
\resizebox{\hsize}{!}{\includegraphics{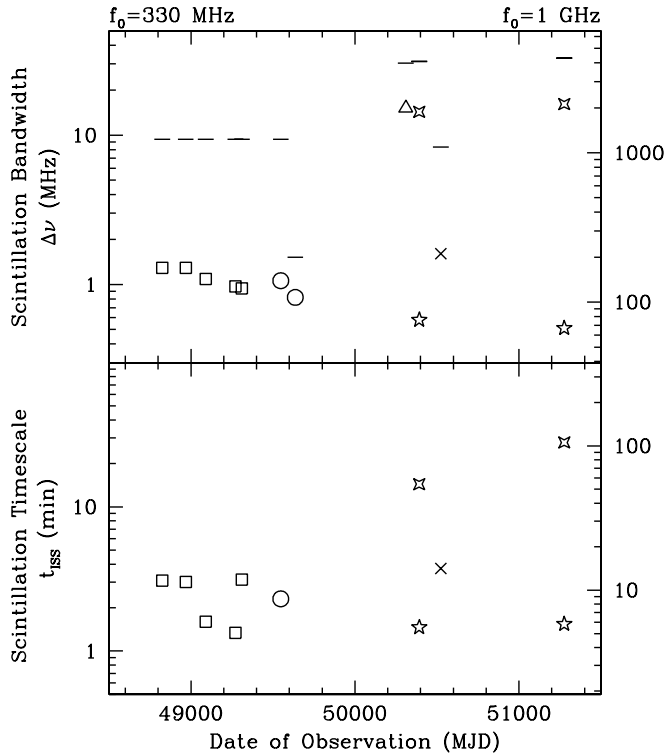}}
\caption{Measurements of the scintillation bandwidth and scintillation
time scale plotted against observing date.  
Results are scaled to $f_0=330$\ MHz (left vertical scale)
and  $f_0=1$\ GHz (right scale) for purposes of comparison.
The horizontal lines
indicate the effective bandwidth used for measurements of $\Delta\nu_{ISS}$.
Symbols show reference:
squares: Nicastro \& Johnston (\cite{nic95});
circles: Johnston et al. (\cite{joh98});
triangle: Issur (\cite{iss00});
stars Paper 1 (4-pointed: broad scintillation, 5-pointed: narrow);
cross Gothoskar \& Gupta (\cite{got00}).
\label{mjd}}
\end{figure}

Table\ \ref{obstable} compares measurement of scintillation timescale and bandwidth
for PSR\ J0435$-$4715, 
from the literature.
For easier comparison, we scaled these measurements to 
a single observing frequency
$f=330$\ MHz,
using the 
scaling relations appropriate for a Kolmogorov spectrum, $n=11/3$:
$t_{ISS}\propto f^{1.2}$ and
$\Delta\nu_{ISS}\propto f^{4.4}$.
Fig.\ \ref{mjd} shows the same data, in graphical form.
Column 7 of Table\ \ref{obstable} gives the observing bandwidth scaled to 330\ MHz,
corresponding to the horizontal lines shown in Fig.\ \ref{mjd}.
The measurements
fall into two clear groups: those showing narrow-band scintillations with $\Delta\nu \approx 1$\ MHz, and
those showing broader-band scintillation with $\Delta\nu\approx 15$\ MHz.  
Observations cannot determine scintillation bandwidth if greater than the observing
bandwidth $B$.
Fig.\ \ref{mjd} shows this upper limit for each observation.
As the figure shows, Issur (\cite{iss00}) and Paper 1
had sufficient bandwidth to detect the wider-band scintillations;
other observations could detect only the narrow-band scintillations.

The scintillation bandwidth of the narrow-band scintillation
varies between epochs,
as is found for some other nearby pulsars (Bhat et al. \cite{Bha99}).
For PSR J0437$-$4715 this variability is most pronounced in the observations
of Gothoskar \& Gupta (\cite{got00}),
who found that the measured $\Delta\nu_{ISS}$ spans a range of more than an
order of magnitude within 2 days.
Paper 1 finds a value for the narrow-band scintillation
at both epochs that
lies within this range;
Gothoskar et al. observed between their two epochs.
Nicastro \& Johnston (\cite{nic95}) and Johnston et al. (\cite{joh98})
observed variation of only
30\% over more than 2 years.
The ranges of values the two groups measure for the narrower $\Delta\nu_{ISS}$
overlap, and have similar centroids.
We will show in \S\ref{inferred_mod_index} and \S\ref{results} that all observed scales of scintillation can be explained
by one power-law spectrum of inhomogeneities.

\section{Structure Function}

\subsection{Construction of the Structure Function}

Only a few of the observations of PSR J0437$-$4715 provide the shape of the correlation function,
which can be used to construct the phase structure function $D_s(t,f)$.
As Shishov et al. (\cite{shi03}) showed, 
$D_s(\Delta t)$ can be obtained for small time lags $\Delta t$
from the correlation function $B_I(\Delta t)$ of intensity variations:
\begin{equation}
D_s(\Delta t)={{B_I(0)-B_I(\Delta t)}\over{\langle I\rangle^2}}
\quad{\rm for\ }\Delta t \leq t_{ISS},
\label{eq1}
\end{equation}
where $t_{ISS}$ is the characteristic time scale of variations caused by electron density
inhomogeneities in the ISM.
In the frequency domain we have:
\begin{equation}
D_s(\Delta f)={{B_I(0)-B_I(\Delta f)}\over{\langle I\rangle^2}}
\quad{\rm for\ }\Delta f \leq \Delta\nu_{ISS},
\end{equation}
where $\Delta\nu_{ISS}$ is the characteristic frequency scale of scintillation, and $\Delta f$ is a frequency difference.

To scale all data to a single reference frequency $f_0$,
we must rescale the structure functions at frequency $f$ appropriately:
\begin{equation}
D_s(f_0,\Delta t(f_0),\Delta f(f_0))=D_s(f,\Delta t,\Delta f) (f_0/f)^2 .
\label{eqfscale}
\end{equation}
The time difference, $\Delta t$, characterizes the same spatial scale independently
of observing frequency, $\Delta t(f_0)=\Delta t(f)$.
However,
$\Delta f(f)\neq \Delta f(f_0)$. 
This is because
wave scattering
by inhomogeneities with a spatial scale $b$ determines the decorrelation of intensity fluctuations at 
frequency difference $\Delta f$. 
Moreover, the relation between $b$ and $\Delta f$ depends on observing frequency $f$.
One may consider two different models for the relation between $b$ and $\Delta f$: diffractive and refractive
(Shishov et al. \cite{shi03}). 

In the diffractive model, decorrelation with frequency arises from changes
in the scattering angle with frequency. The frequency difference $\Delta f_d$ corresponding to the scale $b$ is
\begin{equation}
\Delta f_d \approx c(kb)^2/R \propto b^2 f^2,
\label{eq4}
\end{equation}
where $1/(kb)$ is the typical scattering angle $\theta_{sc}$, 
$k$ is the wave number and $R$ is an effective distance
to the scattering layer (Ostashev \& Shishov \cite{ost77}, Shishov et al. \cite{shi03}).
Thus, for diffractive scintillation, one obtains:
\begin{equation}
\Delta f_d(f_0)=(f_0/f)^2 \Delta f(f) .
\label{eq5}
\end{equation}

In a model including strong angular refraction, another relation
between $\Delta f$ and $b$ is realized. We 
introduce the typical refractive angle $\theta_{ref}$. 
If $\theta_{ref} \gg 1/(kb)$, the displacement of a beam path with
changing frequency converts the spatial pattern to structure in frequency.  
The frequency difference, $\Delta f_r$,
corresponding to the scale $b$ is then (Shishov et al. \cite{shi03}):
\begin{equation}
\Delta f_r \approx c(kb)/(R\theta_{ref}) \propto b f^3 .
\label{eq6}
\end{equation}
So for the case of strong angular refraction:
\begin{equation}
\Delta f_r (f_0)=(f_0/f)^3 \Delta f(f) .
\label{eq7}
\end{equation}
We will use these relations in our analysis of the data.

\subsection{Inferred Scintillation Parameters of PSR\ J0437$-$4715}

As shown in detail in Table\ \ref{obstable},
scintillation parameters for PSR\ J0437$-$4715
obtained at different frequencies and different times differ very strongly
even when referred to a single frequency: by about a factor of 30.
We observe that there can be at least two explanations for such behavior.
One is that multiple scales of scintillation are present;
one might expect such a model in the interstellar Levy flight proposed by 
Boldyrev \& Gwinn (\cite{bol03}),
for example.
In  this case we might see fine frequency structure caused by occasional large-angle deflections,
leading to large propagation times, as well as the coarser structure originating from more typical deflections.

An alternative explanation is that we do not observe the
fundamental scales of scintillation:
for example, if the bandwidth of the receiver is less than the actual diffractive scale,
then one observes only the tail of the fast variations of intensity in frequency and
time domains.  This tail has modulation index $m << 1$.
Usually observers do not record the modulation index,
but rather only the scintillation bandwidth and timescale
as described above.  
When the bandwidth of the analysis is less than the diffractive scale of scintillation,
$\Delta\nu_{ISS}$,
these scales do not correspond to the actual time or frequency scale of scintillation.

\subsection{Inferred Modulation Index}\label{inferred_mod_index}

Only two observations of those listed in Table\ \ref{obstable} had sufficiently wide observed bandwidth
to detect the fundamental scale of scintillation:
Issur (\cite{iss00}) and Paper 1.
All other observers had bandwidths several times less than
the characteristic frequency scale of scintillation.
Unfortunately they did not publish observed values of the modulation index,
but we can determine the expected modulation index theoretically if
we know the ratio of the observing bandwidth $B$ to the scintillation bandwidth $\Delta\nu_{ISS}$,
and the ratio of the time span of a scan $T$ to the scintillation timescale $t_{ISS}$.

As is shown in the Appendix, the modulation index is expected to be defined by the largest
of the ratios of these parameters, if both ratios are less than one.
We note that the modulation index will be different for diffractive and refractive models (see Appendix).
In the presence of strong angular refraction, when the refraction angle is much more than the scattering angle, 
$\theta_{ref}>>\theta_{dif}$, a refractive model is required (Shishov et al. \cite{shi03}).

We determined the expected modulation indices from the observing bandwidth and time span,
and show the results in Table\ \ref{modtable}.
For all of the observations, $T/t_{ISS}>1$,
but for some we have $B/\Delta\nu_{ISS}<1$. 
In this case, the expected modulation index
is given by (see Appendix):
\begin{eqnarray}
m_{ed}^2&=&\left[{{2.8}\over{(n(n+2))}}\right](B/\Delta\nu_{ISS})^{(n-2)/2},\quad{\rm diffractive\ model,} \label{modexpect8}\\
m_{er}^2&=&\left[{{0.7}\over{(n(n-1))}}\right](B/\Delta\nu_{ISS})^{(n-2)},\quad{\rm refractive\ model.} \label{modexpect9}
\end{eqnarray}
Here, $n$ is the power-law index of the spectrum of density fluctuations in the ISM: for example,
for the Kolmogorov theory $n=11/3$.
In the standard calculation of 2D (frequency and time) autocorrelation functions of scintillation spectra,
and determination of $t_{ISS}$ and $\Delta\nu_{ISS}$ from them, 
the mean intensity from the observations
is subtracted from the intensity data before autocorrelation
(Cordes \cite{cor86}).
Equation \ref{modexpect8} or \ref{modexpect9} holds in this case.
Table\ \ref{modtable} includes the expected value of the modulation index
for both models, $m_{ed}$ and $m_{er}$.
These values were calculated from the expected values of the scintillation bandwidth $\Delta\nu^0$
and the timescale $t_{ISS}^0$.
We found these expected values from the values reported by Paper 1:
$\Delta\nu_{ISS}=16$\ MHz and $t_{ISS}=17$\ minutes for $f=328$\ MHz. 
We scaled these values to the observing
frequency $f$ using the scaling laws appropriate for a Kolmogorov spectrum, $n=11/3$:
$\Delta\nu^0\propto f^{4.4}$ and $t_{ISS}\propto f^{1.2}$.
For the refractive model, the scaling law is actually $\Delta\nu^0\propto f^{4.2}$, but this does not change our
estimates by more than 10\%.

We note that the frequency scale measured at 152\ MHz agrees well with that extrapolated from
328\ MHz, but the estimate of the decorrelation time has a large error bar because the ratio $T/t_{ISS}$ is small.
In this case, it is better to use $t_{ISS}$ defined at the level of 0.5, rather than 1/e.  We did not include the
values of $t_{ISS}$ and $\Delta\nu_{ISS}$ measured at 660\ MHz
by Johnston et al. (\cite{joh98}) 
because the ratio $B/\Delta\nu_{ISS}$ is very small for those measurements,
so that the mean intensity cannot be defined reliably.

Paper 1 reports two frequency scales: $\Delta \nu_{1}=16$\ MHz and $\Delta\nu_{2}=0.5$\ MHz,
both at an observing frequency of $f=328$\ MHz.
The small-scale scintillation was detectable with a reduced observing bandwidth of $B=8$\ MHz,
as we discuss further below.
They found that the modulation index for this structure is $m\approx 0.2$, which is close to what we would estimate from
Eq. \ref{modexpect9}.
Gothoskar \& Gupta (\cite{got00}) report $\Delta\nu_{ISS}\approx 2.5$\ MHz at the same frequency.  This differs markedly from the value reported
by Paper 1.
Their receiver bandwidth was less than half the expected decorrelation bandwidth,
but they had finer frequency resolution than Paper 1.

The structure function of intensity 
fluctuations with frequency
corresponds to the actual structure function of interstellar scintillation
only when the frequency bandwidth of the observations, $B$, is significantly 
larger than the frequency scale of the scintillations. 
In that case one 
obtains the true characteristic frequency scale. 
In 
the case of limited observing bandwidth $B$, one observes only a tail of the 
actual
frequency structure function, corresponding to the small scales of 
variations. The scales of these variations obtained 
from the autocorrelation function will 
depends on the parameters of the observation, and can be used for construction 
of a composite structure function only by using the proper modulation index (see 
Eq.\ \ref{eq10}).  
In the case of insufficient bandwidth and time span of observation, 
the scintillation bandwidth and scintillation timescale obtained 
from the autocorrelation function 
can differ strongly from the actual ones.

\section{Results}\label{results}

We constructed a composite time and frequency structure function using correlation functions at different frequencies,
and converting them to a frequency $f_0$, using Eqs.\ \ref{eq1}-\ref{eqfscale} and either \ref{modexpect8} or 
\ref{modexpect9}, and either \ref{eq5} or \ref{eq7} for the frequency correlation function. 
We used the frequency correlation function obtained by Issur at 152\ MHz,
frequency and time correlation functions at $f=327$\ MHz from the paper of 
Gothoskar \& Gupta (\cite{got00}),
and data at 328\ MHz of Paper 1.
We normalized the correlation functions by dividing them by their values at zero lag, so all are normalized at zero lag. We did so
because we do not know the actual values of mean and rms intensity for the different observations.
In this case,
\begin{equation}
D_s(\Delta t,\Delta f)=(1-B(\Delta t,\Delta f)) m^2 .
\label{eq10}
\end{equation}
We chose to use $f_0 = 1000$ MHz, 
to simplify comparison with structure functions 
for other pulsars determined
by Shishov et al. (\cite{shi03}) and Smirnova et al. (\cite{smi06}).

In Fig.\ \ref{fig1}, we present the phase structure function.
Note that the data from Gothoskar \& Gupta (\cite{got00}) and Paper 1 agree very well.
The expected modulation index, as calculated in Table\ \ref{modtable},
has been used in setting the overall offset of the data of Gothoskar \& Gupta.
We have estimated the value of the structure function at 436\ MHz based
on the characteristic timescale reported by Johnston et al. (\cite{joh98}),
and associating that point with the expected modulation index given in 
Table\ \ref{modtable},
for the refractive model.
This point is in good agreement with the other points,
and that good agreement leads us to the conclusion that all spatial scales of
inhomogeneities probed in these observations arise from a single spatial  spectrum.
A fit to all points gives us the slope $\alpha = 1.46\pm 0.20$,
which corresponds to $n=3.46\pm 0.20$.
To convert this structure function of time lag $\Delta t$ to a function of spatial scale $\ell$,
we use the simple conversion
as $\ell = \Delta t\, V_{ISS}$, where $V_{ISS}$ is the speed of the diffraction pattern
relative to the observer.
To choose $V_{ISS}$ correctly, we must know the location of the scattering material along the line of sight.
We consider this location and the appropriate conversion in \S\ref{discussion} below.

In Fig.\ \ref{fig2}, we show the frequency structure functions for two models:
a model with a strong angular refraction (upper panel) and a diffractive model (lower panel).
The data have been combined using Eqs.\ \ref{eq1} - \ref{modexpect9}.
All of the data agree well for both models, but the slope of the structure function 
($\alpha = 1.44 \pm 0.03$)
is about the same as for the temporal structure function. This leads to the conclusions that we have a
refractive model: for the diffractive model the slope should rather be half that of the temporal
structure function (Shishov et al. \cite{shi03}).

\section{Discussion}\label{discussion}

Several components of the ISM are responsible for
the interstellar scintillation of pulsars and extragalactic radio sources.
Component A is homogeneously distributed up to a distance of order 1\ kpc from the Sun.,
and is distributed in the space outside the spiral arms of the Galaxy
(Cordes \cite{cor85}, Cordes et al. \cite{cor91}, Pynzar' \& Shishov \cite{pyn97}).
At larger distances component B, located in the spiral arms of the Galaxy,
gives the primary contribution to scintillation effects.
A third component lies about 10\ pc from  the Sun, where it contributes
an enhanced level of turbulence 
(Jauncey et al. \cite{jau00},
Dennett-Thorpe \& de Bruyn \cite{den02},
Rickett et al. \cite{ric02}).
We refer to this component as component C.
Component C is primarily responsible for the scintillation of extragalactic radio sources, 
because the large angular sizes of the sources suppress the influence of other components.
For the closest pulsars, the scintillation effects caused by components A and C
can be comparable,
and we must investigate especially the relative contribution of these two components in
scintillation, for each pulsar.

Pulsar PSR\ J0437$-$4715 is one of the closest pulsars.
Estimation of the relative contributions of components A and C for this object
depends on the relative
values of the spatial diffractive scale $b$ and the Fresnel scale $b_{Fr}=(R/k)^{1/2}$,
where $k=2\pi/\lambda$ is the wavenumber, and $R$ is the distance to the scattering layer for the
pulsar.

\subsection{Model for Scattering by Component A}

Suppose first that component A provides the primary contribution to the scintillation of
PSR\ J0437$-$4715. Using the distance to the pulsar of $L=150$\ pc 
(van Straten \cite{Van01}),
we obtain the value of the Fresnel scale for component A (here $L \approx R$) at frequency
$f=330$\ MHz, and find $b_{Fr} = (R/k)^{1/2}\approx 10^{11}$\ cm.
Using the pulsar velocity of $V=100$\ km/s (van Straten \cite{Van01}),  
we obtain the characteristic spatial scale of the diffraction pattern
$b_{dif}=V t_{ISS}\approx 10^{10}$\ cm.
Because $b_{Fr}>b_{dif}$, the scintillation must be saturated.
Using the value of $b_{dif}$ we can estimate the scattering angle
$\theta_{sc} \approx 1/(k b_{dif})\approx 0.3$\ mas.
The expression for the scintillation bandwidth can be obtained from 
Equation \ref{eq4}:
\begin{equation}
\Delta\nu^0_{dif}\approx c/(R\theta_{sc}^2) .
\label{eq11}
\end{equation}
Using  Eq\ \ref{eq11} and
the scattering angle $\theta_{sc}=0.3\ {\rm mas}$,
we obtain the estimated scintillation bandwidth for the diffraction pattern: $\Delta\nu^0_{dif}\approx 30\ {\rm MHz}$.
This expected scale is about twice the observed scale $\Delta\nu_{ISS}$.
However, 
in \S\ref{results} above
we show that for PSR\ J0437$-$4715 the
slopes of the time and frequency structure functions
are best described by the refractive model.
The characteristic frequency scale is thus determined by the expression 
(Shishov et al. \cite{shi03}):
\begin{equation}
\Delta\nu^0_{ref} \approx c/(R \theta_{ref}\theta_{sc})
\label{eq12}
\end{equation}
For a refractive angle of, for example, $\theta_{ref} \approx 2\theta_{sc}$ at 
observing frequency 
$f^0 = 330$\ MHz we obtain for the decorrelation bandwidth:
$\Delta\nu^0_{ref} \approx 15$\ MHz.  This value is consistent with the measured value.

However, the value of $\theta_{sc} \approx 0.3$ mas is much smaller than the value of
$\theta_{sc} \approx 2$ mas  estimated from
the statistical dependence of $\theta_{sc}$ on DM, given in the paper 
Pynzar' \& Shishov (\cite{pyn97}) using the dispersion measure of
PSR J0437$-$4715, $DM = 2.65$ pc/cm$^3$ (see Paper 1).
Therefore, Component A explains the observational data only with difficulty.

\subsection{Model for Scattering by Component C}

We now suppose that component C of the interstellar medium makes the greatest contribution
to the scintillation of PSR\ J0437$-$4715. Using a distance of $R=10$\ pc for component C,
at 
$f^0=330$\ MHz we obtain for the Fresnel scale $b_{Fr}=(R/k)^{1/2}\approx 2\times 10^{10}$\ cm.
Using an observer speed of $V=30$\ km/s, dominated by the Sun for this nearby material, we find for the characteristic spatial scale of the
diffraction pattern $b_{dif}=V t_{ISS} \approx 3\times 10^9$\ cm.
Scintillation is again in the saturated regime in this case. Using this value of $b_{dif}$, we estimate
the scattering angle as $\theta_{sc}\approx 1/(k b_{dif}) \approx 1$ mas.
Substituting this value of the scattering angle in Eq.\ \ref{eq11}, 
we obtain the estimated scintillation bandwidth for the diffractive model
$\Delta\nu^0_{dif} \approx 40$ MHz. 
However, as argued above, the time and frequency structure
functions favor a refractive model.
For consistency with the measured value of decorrelation bandwidth 
$\Delta\nu^0_{ref} = 16$\ MHz
at $f^0 = 330$\ MHz,  
we take $\theta_{ref}= 1.5 \theta_{sc}$ and obtain $\Delta\nu^0_{ref} = 15$\ MHz.
In this case, the value of $\theta_{sc} \approx 1$ mas corresponds much better 
than for the Component A model
to the expected value of
$\theta_{sc} \approx 2$ mas for this pulsar.
We therefore conclude that component C can explain the observational data better.

\subsection{Comparison of Scintillation Parameters for PSR\ J0437$-$4715 and PKS\ 0405$-$385}

Quasar PKS\ 0405$-$385 is located about $10^{\circ}$ from PSR\ J0437$-$4715 on the sky.
The interstellar scintillation of this quasar is certainly determined by component C of the
turbulent interstellar plasma
(Dennett-Thorpe \& de Bruyn \cite{den02}, Rickett et al. \cite{ric02}).
The scintillation of PKS\ 0405$-$385 is weak at $f=4.8$\ GHz, with a measured scintillation
index of $m \approx 0.1$ (Rickett et al. \cite{ric02}).
Prokhorov et al. (\cite{pro75}) showed that in weak scintillation,
the square of the modulation index is about equal to the value of the phase structure
function at the Fresnel scale:
\begin{equation}
D_S(b_{Fr})\approx m^2 \approx 0.01 \quad {\rm for\ } f=4.8\ {\rm GHz} .
\end{equation}
For a power-law spectrum with power-law index $n$, the value of $D_S(b_{Fr})$
at frequency $f_0$
can be scaled to the given frequency $f$ by the relation:
\begin{equation}
D_S(f,b_{Fr}(f)) = (f_0/f)^{(n+2)/2} D_S(f_0,b_{Fr}(f_0)) .
\end{equation}
Here we used the frequency dependence of modulation index (Shishov \cite{shi93}).
Using this relation with $n=3.6$ and the above value for the phase structure function at $f=4.8$\ GHz,
we obtain the estimate
$D_S(b_{Fr})\approx 20$ at $f_0=330$\ MHz.
We can also scale the timescale for weak scintillation $t_{w}=33$\ min
at 4.8\ GHz (Rickett et al. \cite{ric02}) to $f_0=330$\ MHz using the relations
\begin{eqnarray}
t_{cr} &\approx &(f_0/f_{cr})^{1/2}  \\
t_{ISS}&\approx&(f_0/f_{cr})t_{cr},\quad f_0<f_{cr}. \nonumber
\end{eqnarray}
Here $f_{cr}$ is the frequency that divides weak and strong scintillation regimes.  It is given
by the equation
\begin{equation}
D_S(f_{cr},b_{Fr}(f_{cr}))=(f_0/f_{cr})^{(n+2)/2} D_S(f_0,b_{Fr}(f_0))=1.
\end{equation}
Using this equation and the fact that $D_S(f,b_{Fr}(f))=0.01$ at $f=4.8$\ GHz,
we obtain $f_{cr}=1$\ GHz, $t_{cr}=70$\ min.
The predicted time scale of $t_{ISS}=23$\ min at frequency $f_0=330$\ MHz is
close to the value obtained for PSR\ J0437$-$4715. This suggests that
the same scattering medium causes the flux
variations of quasar PKS\ 0405$-$385 and the scintillation of PSR\ J0437$-$4715. 

\begin{figure}
\resizebox{\hsize}{!}{\includegraphics{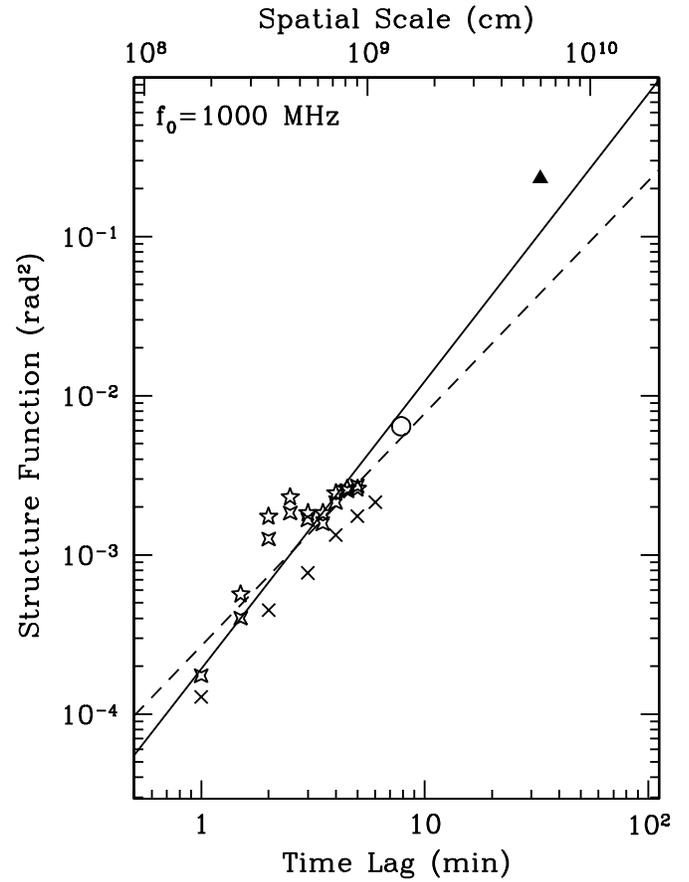}}
\caption{The time structure function of phase fluctuations
for PSR\ J0437$-$4715
reduced to the reference frequency $f_0 = 1000$\ MHz,
as compiled from the observations.
Open symbols are as in Figure 1.
The solid triangle indicates scintillation of the quasar PKS\ 0405$-$385 
Rickett et al. (\cite{ric02}).
The dashed line indicates the best fit to data for the pulsar (index $\alpha_1=1.46\pm 0.20$);
the solid line corresponds to the best-fitting power-law to all points (index $\alpha_2=1.8\pm 0.15$). 
Top x-axis corresponds to spatial scale of the inhomogeneities.
\label{fig1} }
\end{figure}

Another, stronger indication
of this can be obtained by including the point corresponding to 
quasar scintillation parameters in Fig. \ref{fig1} (triangle).
Here we use a time scale of 33\ min and the value of $D_S = 0.01$ scaled to $f_0 = 1$ GHz. 
The point is
in a good agreement with extrapolation from the pulsar data. 
A fit to all points gives the slope $\alpha = 1.8 \pm 0.15$ (the solid line
in Fig. 2).
Using this model of the local
scattering layer (with $V = 30$ km/s) we can reduce
the time and frequency structure functions to a spatial form.
Fig.\ \ref{fig1} (top x axis) shows the inferred phase structure function, using conversion of the of
time scale to
spatial scale as $b = 3\times 10^6\ {\rm cm\ s}^{-1} \Delta t$.

\section{Conclusions}

Observers have reported scintillation bandwidths for PSR\ J0437$-$4715 
that range by a factor of about 30, when scaled to a single observing 
frequency.
The observations of Paper 1 found two scales, at each of two observations.
In this paper, we show that all of those observations can be represented by a single power-law
structure function,
when corrections for incomplete sampling in time and frequency are applied.
Results of broad-band observations correspond to the actual scales of 
interstellar scintillation, including the characteristic scale at 
the observing frequency.
Results of narrow band observations correspond to the actual interstellar
scintillations only for small frequency lags, $\Delta f << B$.  Thus, in
many cases, the observational papers presented only part of
the information carried by the structure function; in these cases we reconstructed the missing information
via theoretical estimates.  
Scintillation bandwidth and scintillation timescale represent only two parameters from the 
extensive information carried by the autocorrelation function;
indeed, the smaller of the two scintillation bandwidths found at each observing epoch in Paper 1 were 
obtained by restricting the bandwidth to a smaller range.

Additional observations, that record modulation index at the central peak and
the functional form of the structure function,
can test our suggestion that all scintillations result from a single power-law structure function.
Alternatively, the wide range of scales might reflect an interstellar Levy flight as proposed by
Boldyrev \& Gwinn (\cite{bol03}).
We respectfully request that observers preserve a greater portion of the
rich information carried by the 
autocorrelation function of intensity in future observations.

Further analysis of the scintillation data for PSR\ J0437$-$4715 indicates that they can explained by a model
for a layer with enhanced turbulence, at distance of about 10\ pc from the Sun.
This model appears to work better than models for scattering by more distant material.
The flux variations of the extragalactic source PKS\ 0405$-$385 located nearby pulsar are consistent with
scattering by material at the same distance and with the same scattering strength.

\begin{acknowledgements}
We thank the U.S. National Science Foundation for supporting this collaboration.
This work was supported by NSF grant No. AST
0098685,the Russian Foundation for Basic Research, project codes
03-02-509, 03-02-16522.
\end{acknowledgements}

{
\appendix
\section{Structure Function for Limited Data}

Estimation of the structure function of intensity fluctuations 
has peculiar aspects, for observations limited in time span $T$
or bandwidth $B$.
The main problem lies in proper normalization of the correlation function.
For observations with limited $T$ and $B$, the correlation function
calculated from observations is
\begin{eqnarray}
C_I(\Delta t,\Delta f) &=& B_I(\Delta t,\Delta f)/B_0, \\
&&{\rm where\ } B_0=B_I(\Delta t=0,\Delta f=0) . \nonumber
\end{eqnarray}
However, the desired correlation function should be normalized by the square of the ensemble-averaged intensity:
\begin{equation}
K_I(\Delta t,\Delta f) = B_I(\Delta t,\Delta f)/\langle I\rangle_{ens}^2.
\end{equation}
Here, the subscripted angular brackets $\langle ... \rangle_{ens}$ indicate an ensemble average.
The corresponding conversion factor is the square of the modulation index:
\begin{equation}
m_f^2 = B_0/\langle I\rangle_{ens} ^2 .
\end{equation}
The subscript ``$f$" on $m$ indicates averaging in the frequency domain;
analogously, ``$t$" denotes the time domain.

For values of the time interval $T$, or frequency bandwidth $B$, that are
large in comparison with the characteristic time and frequency scales of scintillation $t_{ISS}$ and $\Delta\nu_{ISS}$,
the modulation index $m_f^2=1$ and $C_I=K_I$.
For small values of $(T/t_{ISS})$ or $(B/\Delta\nu_{ISS})$,
the intensity fluctuations along these dimensions
are intrinsically non-stationary stochastic processes,
the variance depends on the averaging interval,
and the estimated scintillation index is less than one.

In the case where the averaging procedure corresponds to integration in the frequency domain,
the variance of the intensity fluctuations is determined by the equations:
\begin{eqnarray}
\langle (\Delta I)^2\rangle_f = (1/B) \int_0^B df' \left[ I(f') - \langle I\rangle_f \right]^2 ,
\quad {\rm where} \label{4A} \\
\langle I \rangle_f = (1/B) \int_0^B df' I(f')  .
\end{eqnarray}
Here the angular brackets with subscript ``$f$'' denote integration over frequency as the averaging
procedure.  The variance (\ref{4A}) is a random value.
Forming the ensemble average, we find
\begin{eqnarray}
&& \\
\langle \langle(\Delta I)^2\rangle_f \rangle_{ens}
&=& {{1}\over{B^2}} \int_0^B df_1 \int_0^B df_2
\left[ \langle I^2\rangle_{ens} - \langle I(f_1) I(f_2) \rangle_{ens}\right]  \nonumber \\
&=& {{1}\over{B^2}} \int_0^B df_1 \int_0^B df_2
\left[ B_{I,ens}(0) - B_{I,ens}(f_1-f_2) \right],  \nonumber
\end{eqnarray}
where $B_{I,ens}(f_1-f_2)$ is the intensity correlation function averaged over the ensemble.
We have added the subscript ``$ens$'' to specifically denote the ensemble averaging.
For a refractive or diffractive model forming the frequency structure of the diffractive scintillation,
the correlation function can be presented by the expression (Shishov \cite{shi03}):
\begin{equation}
B_{I,ens}(f_1-f_2) = \langle I\rangle_{ens}^2  \exp[-D_S(f_1-f_2)]
\end{equation}
The dependence of $D_S(f_1-f_2)$ on $(f_1-f_2)$
is different for refractive and diffractive models.
For the diffractive model, we have (Shishov \cite{shi03}):
\begin{equation}
D_S(f_1-f_2) = 0.7 [(f_1-f_2)/\Delta\nu_{ISS}]^{(n-2)/2} .
\end{equation}
Using this function we obtain
\begin{equation}
\langle m_f^2\rangle = [2.8/(n(n+2))] (B/\Delta\nu_{ISS})^{(n-2)/2} . \label{mf_diffract}
\end{equation}
For the refractive model we have:
\begin{equation}
D_S(f_1-f_2) = 0.7 [(f_1-f_2)/\Delta\nu_{ISS}]^{(n-2)} .
\end{equation}
Using this function we obtain
\begin{equation}
\langle m_f^2\rangle = [0.7/(n(n-1))] (B/\Delta\nu_{ISS})^{(n-2)} .  \label{mf_refract}
\end{equation}

The measured value of the scintillation index is a partly averaged one,
and in the case of small values of $(T/t_{ISS})$ or $(B/\Delta\nu_{ISS})$ it can fluctuate,
by about 100\%.

From similar considerations for the temporal version of the averaging procedure
we obtain
\begin{equation}
\langle m_t^2\rangle = [1/(n(n-1))] (T/t_{ISS})^{(n-2)} .  \label{mt}
\end{equation}

If we employ two-dimensional integration, in frequency and time,
the scintillation index is determined by either Eq.\ \ref{mf_diffract} or \ref{mf_refract}
for $(T/t_{ISS})<<(B/\Delta\nu_{ISS})$, and by Eq.\ \ref{mt} for the reverse case.

}

\clearpage

% mjd/mjd_redodata.ps

% Russia_Fig2/RFig2.ps

%  Russia_Fig3/RFig3.ps
\begin{figure}
\resizebox{\hsize}{!}{\includegraphics{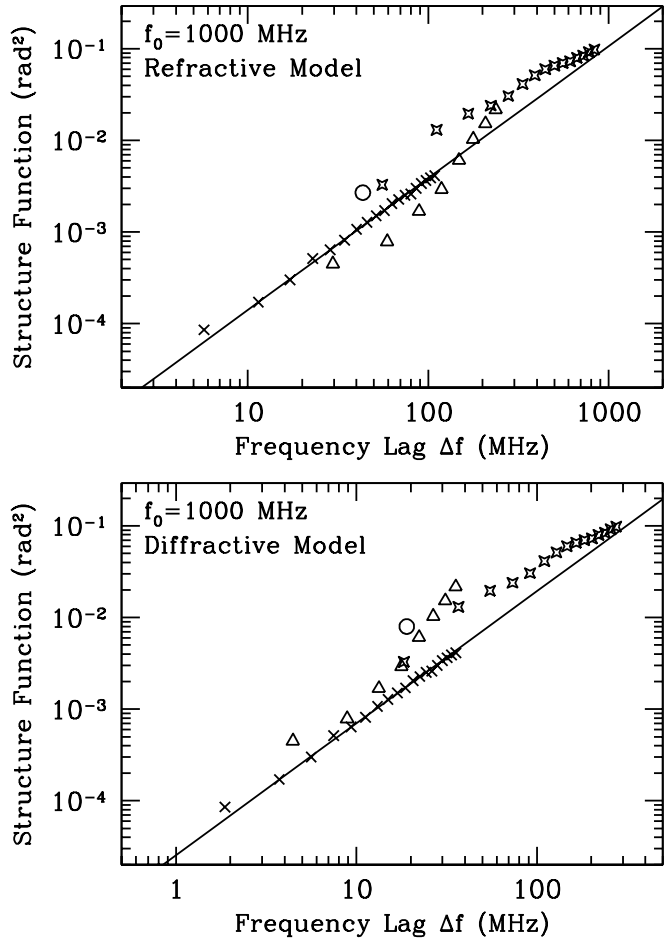}}
\caption{The frequency structure function of phase fluctuations,
reduced to the reference frequency $f_0 = 1000$\ MHz,
as compiled from the observations, using the same symbols as in Figure 1.
The solid lines show the diffractive model (upper) and the refractive model (lower).
Slopes were fit to the data taken at $f=327$\ MHz.
\label{fig2}}
\end{figure}

\clearpage

\begin{table}
\caption{
Measurements of Scintillation Parameters for J0437$-$4715
}
\label{obstable}

\begin{tabular}{lccccccr}
\hline\hline

       &                  &                 &Measured      &Measured     &Scintillation  &Observing &\\
       &Observing  &Observing &Scintillation &Scintillation   &Bandwidth                 &Bandwidth &\\
       &Frequency &Bandwidth &Bandwidth     &Timescale     &scaled to                    &scaled to &\\
Epochs &$\nu$   &          &$\Delta\nu_{ISS}$      &$t_{ISS}$ &330 MHz$^a$               &330 MHz$^a$ &Ref.\\
(MJD)  &(MHz)   &(MHz)     &(MHz)         &(min)         &(MHz)         &(MHz)     &\\
\hline
48825 to 49313 & 436 & 32 & 3.2 to 4.4   & 4.6 to 10.9 &   0.9 to 1.3 & 9.4  & 1 \\
49459 to 49636 & 436 & 32 & 3.61         & 7.8         &   1.1        & 19.  & 2 \\
49636          & 660 & 32 & 17.4         & 10.         &   0.8        &  1.5 & 2 \\
50310          & 152 &  1 & 0.5          & -           &  15.2        & 30.  & 3 \\
50392 to 51276 & 328 & 32 & 16           & 17         &  16.4        & 33.  & 4 \\
50392 to 51276 & 328 &  8 & 0.5          & 1.5         &   0.5        & 8.  & 4 \\
50523 to 50524 & 327 &  8 & 0.18 to 2.96 & 1.8 to 5.0  &  0.19 to 3.1 & 8.3  & 5 \\
\hline
\end{tabular}

\par\noindent
$^{a}$\ {Scaled to $\nu=330$ MHz using $\Delta\nu \propto \nu^{4.4}$.}

\par\noindent 
References: 
(1) Nicastro \& Johnston 1995; 
(2) Johnston, Nicastro, \& Koribalski 1998;
(3) Issur 2000;
(4) Gwinn et al. 2005 (Paper 1);
(5) Gothoskar \& Gupta 2000.

\end{table}

\begin{table}
\caption{
Expected Modulation Indices for Observations of PSR J0437$-$4715
}
\label{modtable}
\begin{tabular}{ccccccccccc}
\hline\hline

                 &                 &        &Measured&Measured   &Expected   &Expected   &\multicolumn{2}{c}{Expected}                           & \\
Observing &Observing&Scan&Scintillation&Scintillation&Scintillation&Scintillation &\multicolumn{2}{c}{Modulation Index:}                & \\
Frequency&Bandwidth&Time&Bandwidth &Timescale &Bandwidth  &Timescale  &Diffractive &Refractive &Ref. \\
$f$             &$B$          &$T$  &$\Delta\nu_{ISS}$      &$t_{ISS}$        &$\Delta\nu^0$   &$t_{ISS}^0$& $m_{ed}$            &$m_{er}$            & \\
(MHz)       &(MHz)       &(min)&(MHz)        &(min)         &(MHz)        &(min)          &                                             &                                               & \\

\hline
152  &  1 & 60 &  0.5   & -   & 0.54  & 6.7  & 1  & 1  & 1 \\
327  &  8 & 10 &  2.5  & 5  & 16  & 16.7  & 0.28  & 0.15  & 2 \\
328  & 32 & 44 & 16  & 17  & 16  & 16.7  & 1  & 1  & 3 \\
436  & 32 & 30 & 3.6  & 7.8  & 56  & 23.5  & 0.29  & 0.17  & 4 \\
\hline
\end{tabular}

\par\noindent 
References: 
(1) Issur 2000;
(2) Gothoskar \& Gupta 2000;
(3) Gwinn et al. 2005 (Paper 1);
(4) Johnston, Nicastro, \& Koribalski 1998.

\end{table}

\end{document}